\documentclass[letter,twocolumn,showpacs,preprintnumbers,amsmath,amssymb,nofootinbib]{revtex4}

\usepackage[english]{babel}
\usepackage{graphicx}
\usepackage{dcolumn}
\usepackage{bm}
\usepackage{verbatim}

\newcommand{\ket}[1]{\left| {#1} \right\rangle}
\newcommand{\bra}[1]{\left\langle {#1} \right|}
\newcommand{\ematriz}[3]{\left\langle {#1} \left|{#2}\right|{#3}\right\rangle}	
\newcommand{\braket}[2]{\left\langle {#1}\left|{#2}\right.\right\rangle}
\newcommand{\proj}[2]{\left| {#1} \right\rangle\!\left\langle {#2} \right|}

\newcommand{\biket}[2]{\left| {#1} \right\rangle_{I}\left| {#2} \right\rangle_{IV}}
\newcommand{\pa}{\uparrow\downarrow}
\newcommand{\tr}{\operatorname{Tr}}

\def\slashchar#1{\setbox0=\hbox{$#1$} 
\dimen0=\wd0 
\setbox1=\hbox{/} \dimen1=\wd1 
\ifdim\dimen0>\dimen1 
\rlap{\hbox to \dimen0{\hfil/\hfil}} 
#1 
\else 
\rlap{\hbox to \dimen1{\hfil$#1$\hfil}} 
/ 
\fi}

\begin{document}

\title{Spin and occupation number entanglement of Dirac fields for non-inertial observers}

\author{Juan Le\'on}
\email{leon@imaff.cfmac.csic.es}
\homepage{http://www.imaff.csic.es/pcc/QUINFOG/}

\author{Eduardo Mart\'{i}n-Mart\'{i}nez}%
 \email{martin@imaff.cfmac.csic.es}
\homepage{http://www.imaff.csic.es/pcc/QUINFOG/}
\affiliation{%
Instituto de F\'{i}sica Fundamental, CSIC\\
Serrano 113-B, 28006 Madrid, Spain.\\
}%


\date{\today}

\begin{abstract}
We investigate the Unruh effect on entanglement taking into account the spin degree of freedom of the Dirac field. We analyze spin Bell states in this setting, obtaining their entanglement dependence on the acceleration of one of the partners. Then, we consider simple analogs to the occupation number entangled state $\left|00\right\rangle+\left|11\right\rangle$ but with spin quantum numbers for $\left|11\right\rangle$. We show that, despite their apparent similitude, while the spinless case is always qubit$\times$qubit, for the spin case acceleration produces a qubit$\times$qu4it state. We also introduce a procedure to consistently erase the spin information from our setting preserving occupation numbers. We show how the maximally entangled state for occupation number emerges from our setting. We as well analyze its entanglement dependence on acceleration, obtaining a greater entanglement degradation than in the spinless case.
\end{abstract}

\maketitle

\section{\label{Intro}Introduction}

Despite their apparently separated application areas, general relativity and quantum information are not disjoint research fields. On the contrary, following the pioneering work of Alsing and Milburn \cite{Alsingtelep} a wealth of works \cite{TeraUeda2,ShiYu,Alicefalls,AlsingSchul,SchExpandingspace,Adeschul,KBr,LingHeZ,ManSchullBlack,PanBlackHoles,Steeg} has considered different situations in which entanglement was studied in a general relativistic setting, for instance, quantum information tasks in the proximity of black holes \cite{TeraUedaSpd,PanBlackHoles,ManSchullBlack,Adeschul}, entanglement in an expanding universe \cite{SchExpandingspace,Steeg}, entanglement with non-inertial partners \cite{Alicefalls,AlsingSchul,TeraUeda2,KBr} etc.

Entanglement behavior in non-inertial frames was first considered in \cite{Alsingtelep} where the fidelity of teleportation between relative accelerated partners was analyzed. After this, occupation number entanglement degradation of scalar \cite{Alicefalls} and Dirac \cite{AlsingSchul} fields due to Unruh effect \cite{Unruh,Crispino} was shown. Recent works studied the effect of the instantaneous Wigner rotations and Thomas spin precession on entanglement \cite{AlsingWigner},\cite{AlsingWignerFot}.

The previous work \cite{AlsingSchul} on Unruh effect for Dirac field mode entanglement does not consider the spin of the parties. Hence, only two occupation numbers $n=(0,1)$ are allowed for each mode. Higher values of $n$ are forbidden by Pauli exclusion principle. However, addressing the effect of Unruh decoherence on spin entanglement can only be done by incorporating the spin of the parties in the framework from the very beginning. As a consequence, occupation number $n=2$ is also allowed. This fact will affect occupation number entanglement which has to be reconsidered in this new setting. For this purpose, in this work we will study the case of two parties (Alice and Rob) sharing a general superposition of Dirac vacuum and all the possible one particle spin states for both Alice and Rob. Alice is in an inertial frame while Rob undergoes a constant acceleration $a$.

We will show that Rob --when he is accelerated respect to an inertial observer of the Dirac vacuum-- would observe a thermal distribution of fermionic spin $1/2$ particles due to Unruh effect \cite{Unruh}. Next, we will consider that Alice and Rob share spin Bell states in a Minkowski frame. Then, the case in which Alice and Rob share a superposition of the Dirac vacuum and a specific one particle state in a maximally entangled combination. In both cases we analyze the entanglement and mutual information in terms of Rob's acceleration $a$.

Finally, we will study the case when the information about spin is erased from our setting by partial tracing, remaining only the occupation number information. Here, entanglement is more degraded than in \cite{AlsingSchul}. This comes about because more accessible levels of occupation number are allowed, so our system has a broader margin to become degraded.

This paper is structured in the following sections.

In sections \ref{sec2}, and \ref{sec3} we introduce the basic formalism and notation to deal with Dirac fields from the point of view of an accelerated observer taking its spin structure into account. In section \ref{sec4} we study how the Minkowski vacuum state is expressed by an accelerated observer when the spin of each mode is included in the setting, discussing the implications of the single-mode approximation often carried out in the literature \cite{AlsingSchul,AlsingMcmhMil}. Also, we build the one particle state with spin $s$ in Rindler coordinates and analyze the Unruh effect when the spin structure is included. Here we discuss the necessity of tracing over Rindler's region IV for Rob, since it is causally disconnected from region I in which we consider Rob's location. In section \ref{sec5} we analyze how entanglement of spin Bell states is degraded due to Unruh effect. We show that, even in the limit of $a\rightarrow\infty$, some degree of entanglement is preserved due to Pauli exclusion principle. Then we analyze Unruh effect on a completely different class of maximally entangled states (like $\ket{00}+\ket{ss'}$ where $s$ and $s'$ are $z$ component of spin labels) comparing it with the spin Bell states. In section \ref{sec6} we show that the erasure of spin information, in order to investigate occupation number entanglement alone, requires considering total spin states for the bipartite system. Finally, our results and conclusions are summarized in section \ref{sec7}.

\section{The setting}\label{sec2}

We consider a free Dirac field in a Minkowski frame expanded in terms of the positive (particle) and the negative (antiparticle) energy solutions of Dirac equation notated $\psi^+_{k,s}$ and $\psi^-_{k,s}$ respectively:
\begin{equation}\label{field}\psi=\sum_{s}\int d^3k\, (a_{k,s}\psi^+_{k,s}+b_{k,s}^\dagger\psi^-_{k,s})\end{equation}
Here, the subscript $k$ notates momentum which labels the modes of the same energy and $s=\{\uparrow ,\downarrow\}$ is the spin label that indicates spin-up or spin-down along the quantization axis. $a_{k,s}$ and $b_{k,s}$ are respectively the annihilation operators for particles and antiparticles, and satisfy the usual anticommutation relations.

For each mode of frequency $k$ and spin $s$ the positive and negative energy modes have the form
\begin{equation}\label{eq2}\psi^\pm_{k,s} =\frac{1}{\sqrt{2\pi k_0}}u^\pm_s(\bm k) e^{\pm i(\bm k\cdot\bm x- k_0t)}\end{equation}
where $u^\pm_s(\bm k)$ is a spinor satisfying the normalization relations $\pm \bar u^\pm_s(\bm k)u^\pm_{s'}(\bm k)=(k_0/m)\delta_{ss'},\bar u^{\mp}_s(\bm k)u^\pm_{s'}(\bm k)=0$.

The modes are classified as particle or antiparticle respect to $\partial_t$ (Minkowski Killing vector directed to the future). The Minkowski vacuum state is defined by the tensor product of each frequency mode vacuum
\begin{equation}\label{vacua}\ket0=\bigotimes_{k,k'}\ket{0_k}^+\ket{0_{k'}}^-\end{equation}
such that it is annihilated by $a_{k,s}$ and $b_{k,s}$ for all values of $s$.

We will use the same notation as reference \cite{AlsingSchul} where the mode label will be a subscript inside the ket, and the absence of subscript outside the ket indicates a Minkowski Fock state.

In this way, and as a difference with previous works, we will consider the spin structure for each mode, and hence, the maximum occupation number is two. This introduces the following notation
\begin{equation}a^\dagger_{k,s}a^\dagger_{k,s'}\ket0=\ket{ss'_k}\delta_{s,{-s'}}\end{equation}
If $s=s'$ the two particles state is not allowed due to Pauli exclusion principle, so our allowed Minkowski states for each mode of particle/antiparticle are
\begin{equation}\{\ket{0_k}^\pm,\ket{\uparrow_k}^\pm,\ket{\downarrow_k}^\pm,\ket{\uparrow\downarrow_k}^\pm\}.\end{equation}

Consider that we have the following Minkowski bipartite state
\begin{eqnarray}\label{gen1}
\nonumber\ket{\phi_{k_A,k_R}}&=&\mu\ket{0_{k_A}}^+\ket{0_{k_R}}^++\alpha\ket{\uparrow_{k_A}}^+\ket{\uparrow_{k_R}}^++\beta\ket{\uparrow_{k_A}}^+\\*
\nonumber&&\times\ket{\downarrow_{k_R}}^++ \gamma\ket{\downarrow_{k_A}}^+\ket{\uparrow_{k_R}}^++\delta\ket{\downarrow_{k_A}}^+\ket{\downarrow_{k_R}}^+\\*
\end{eqnarray}
with $\mu=\sqrt{1-|\alpha|^2-|\beta|^2-|\gamma|^2-|\delta|^2}$. The subscripts $A,R$ indicate the modes associated with Alice and Rob respectively. All other modes of the field are unoccupied --that is to say that the complete state would be $\ket\Phi=\ket{\phi_{k_A,k_R}}\otimes[\bigotimes_{(k\neq k_A,k_R),k'}\ket{0_k}^+\ket{0_k}^-]$--.

This state generalizes the Bell spin states (for instance, we have $\ket{\phi^+}$ choosing $\alpha=\delta=1/\sqrt2$) or a modes entangled state (for instance choosing $\alpha=\mu=1/\sqrt2$). With this state \eqref{gen1} we will be able to deal with two different and interesting problems at once, 1. Studying the Unruh decoherence of spin entangled states and 2. Investigating the impact of considering the spin structure of the fermion on the occupation number entanglement and its Unruh decoherence.

Later on, under the single mode approximation, we will assume that Alice is stationary and has a detector sensitive only to the mode $k_A$, and Rob moves with uniform acceleration $a$ taking with him a detector sensitive to the mode $k_R$.

\section{Rindler metric and Bogoliubov coefficients for Dirac fields}\label{sec3}

An uniformly accelerated observer viewpoint is described by means of the well-known Rindler coordinates \cite{gravitation}. In order to cover the whole Minkowski space-time, two different set of coordinates are necessary. These sets of coordinates define two causally disconnected regions in Rindler space-time. If we consider that the uniform acceleration $a$ lies on the $z$ axis, the new Rindler coordinates $(t,x,y,z)$ as a function of Minkowski coordinates $(\tilde t,\tilde x,\tilde y,\tilde z)$ are
\begin{equation}\label{Rindlcoordreg1}
a\tilde t=e^{az}\sinh(at),\; a\tilde z=e^{az}\cosh(at),\; \tilde x= x,\; \tilde y= y
\end{equation}
for region I, and
\begin{equation}\label{Rindlcoordreg2}
a\tilde t=-e^{az}\sinh(at),\; a\tilde z=-e^{az}\cosh(at),\; \tilde x= x,\; \tilde y= y
\end{equation}
for region IV.
\begin{figure}\label{fig1}
\includegraphics[width=.45\textwidth]{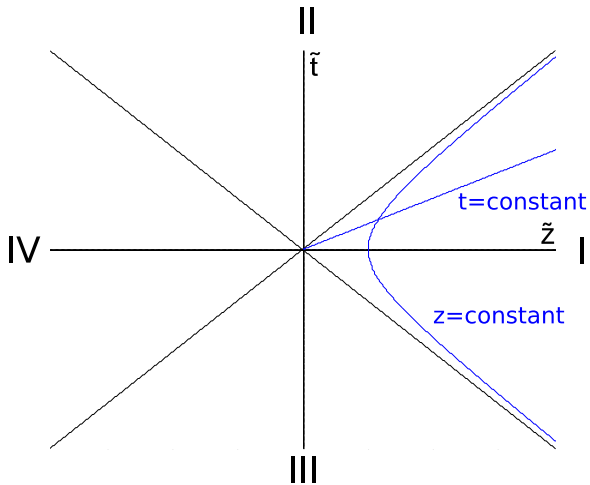}
\caption{Rindler space-time diagram: lines of constant position $z=\text{const.}$ are hyperbolae and all the curves of constant proper time $t$ for the accelerated observer are straight lines that come from the origin. An uniformly accelerated observer Rob travels along a hyperbola constrained to region I}
\end{figure}
As we can see from fig. 1, although we have covered the whole Minkowski space-time with these sets of coordinates, there are two more regions labeled II and III. To map them we would need to switch $\cosh\leftrightarrow\sinh$ in equations \eqref{Rindlcoordreg1},\eqref{Rindlcoordreg2}. In these regions $t$ is a spacelike coordinate and $z$ is a timelike coordinate. However, the solutions of Dirac equation in such regions are not required to discuss entanglement between Alice and an accelerated observer since he would be constrained to either region I or IV, having no possible access to the opposite regions as they are causally disconnected \cite{Birrell,gravitation,Alicefalls,AlsingSchul}.

The Rindler coordinates $z,t$ go from $-\infty$ to $\infty$ independently in regions I and IV. It means that each region admits a separate quantization procedure with their corresponding positive and negative energy solutions\footnote{Throughout this work we will consider that the spin of each mode is in the acceleration direction and, hence, spin will not undergo Thomas precession due to instant Wigner rotations \cite{AlsingSchul,Jauregui}.} $\{\psi^{I+}_{k,s},\psi^{I-}_{k,s}\}$ and $\{\psi^{IV+}_{k,s},\psi^{IV-}_{k,s}\}$.

Particles and antiparticles will be classified with respect to the future-directed timelike Killing vector in each region. In region I the future-directed Killing vector is
\begin{equation}\label{KillingI}
\partial_t^I=\frac{\partial \tilde t}{\partial t}\partial_{\tilde t}+\frac{\partial\tilde z}{\partial t}\partial_{\tilde z}=a(\tilde z\partial_{\tilde t}+\tilde t\partial_{\tilde z}),
\end{equation}
whereas in region IV the future-directed Killing vector is $\partial_t^{IV}=-\partial_t^{I}$.

This means that solutions in region I, having time dependence $\psi_k^{I+}\sim e^{-ik_0t}$ with $k_0>0$, represent positive energy solutions, whereas solutions in region IV, having time dependence $\psi_k^{I+}\sim e^{-ik_0t}$ with $k_0>0$, are actually negative energy solutions since $\partial^{IV}_t$ points to the opposite direction of $\partial_{\tilde t} $ \cite{AlsingSchul,Birrell}. As I and IV are causally disconnected $\psi^{IV\pm}_{k,s}$ and $\psi^{I\pm}_{k,s}$ only have support in their own regions, vanishing outside them.

Let us denote $(c_{I,k,s},c^{\dagger}_{I,k,s})$ the particle annihilation and creation operators in region I and $(d_{I,k,s},d^{\dagger}_{I,k,s})$ the corresponding antiparticle operators. Analogously we define $(c_{IV,k,s},c^{\dagger}_{IV,k,s}, d_{IV,k,s},d_{IV,k,s}^\dagger)$ the particle/antiparticle operators in region IV.

These operators satisfy the usual anticommutation relations $\{c_{\text{R},k,s},c^\dagger_{\text{R}',k',s'}\}=\delta_{\text{R}\text{R}'}\delta_{kk'}\delta_{ss'}$ where the subscript R notates the Rindler region of the operator $\text{R}=\{I,IV\}$. All other anticommutators are zero. That includes the anticommutators between operators in different regions of the Rindler space-time.

Taking this into account we can expand the Dirac field in Rindler coordinates analogously to \eqref{field}:
\begin{eqnarray}\label{fieldri}
\nonumber\psi&=&\sum_{s}\int d^3k\, \left(c_{I,k,s}\psi^{I+}_{k,s}+d_{I,k,s}^\dagger\psi^{I-}_{k,s}+c_{IV,k,s}\psi^{IV+}_{k,s}\right.\\*
&&\left.+d_{IV,k,s}^\dagger\psi^{IV-}_{k,s}\right).\end{eqnarray}

Equations \eqref{field} and \eqref{fieldri} represent the decomposition of the Dirac field in its modes in Minkowski and Rindler coordinates respectively. We can relate Minkowski and Rindler creation and annihilation operators by taking appropriate inner products \cite{Takagi,Jauregui,Birrell,AlsingSchul}. The relationship between Minkowski and Rindler particle/antiparticle operators is linear and the coefficients that relate them are called Bogoliubov coefficients:
\begin{eqnarray}\label{Bogoliubov}
\nonumber a_{k,s}&=&\cos{r}\,c_{I,k,s}-e^{i\phi}\sin r\,d^\dagger_{IV,-k,-s}\\*
b^\dagger_{k,s}&=&\cos{r}\,d^\dagger_{IV,k,s}+e^{-i\phi}\sin r\,c_{I,-k,-s}
\end{eqnarray}
where
\begin{equation}\label{defr}
\tan r=e^{-\pi \frac{k_0c}{a}}
\end{equation}
and $\phi$ is a phase factor that will turn out to be irrelevant for our purposes. Notice that as we are working with two spatial-temporal dimensions and with massless Dirac field, the relation between Rindler modes and Minkowski modes is given in \eqref{Bogoliubov}. We will discuss in the conclusions the implications of considering extra dimensions and massive fields, where Minkowski modes are spread over all positive Rindler frequencies \cite{Takagi}.

Notice from Bogoliubov transformations \eqref{Bogoliubov} that the Minkowski particle annihilator $a_{k,s}$ transforms into a Rindler particle annihilator of momentum $k$ and spin $s$ in region I and an antiparticle creator of momentum $-k$ and spin $-s$ in region IV (in region IV all the magnitudes that are not invariant under time reversal change).

\section{Unruh effect for fermion fields of spin $1/2$}\label{sec4}

Now that we have the relationships between the creation and annihilation operators in Minkowski and Rindler coordinates, we can obtain the expression of the Minkowski vacuum state for each mode $\ket {0_k}$ in Rindler coordinates. For notation simplicity, we will drop the $k$ label in operators/states when it does not give any relevant information, but we will continue writing the spin label.

It is useful to introduce some notation for our states. We will denote with a subscript outside the kets if the mode state is referred to region I or IV of the Rindler space-time. The absence of subscript outside the ket will denote Minkowski coordinates. The $\pm$ label of particle/antiparticle will be omitted throughout the paper because, for the cases considered, a ket referred to Minkowski space-time or Rindler's region I will always denote particle states and a ket referred to region IV will always notate antiparticle states.

Inside the ket we will write the spin state of the modes as follows
\begin{equation}\label{notation1}
\ket{s}_I=c^\dagger_{Is}\ket{0}_I,\qquad \ket{s}_{IV}=d^\dagger_{IVs}\ket{0}_{IV}
\end{equation}
which will notate a particle state in region I and an antiparticle state in region IV respectively, both with spin $s$.

We will use the following definitions for our kets
\begin{eqnarray}\label{notation2}
\nonumber \ket{\uparrow\downarrow}_I&=&c^\dagger_{I\uparrow}c^\dagger_{I\downarrow}\ket{0}_I=-c^\dagger_{I\downarrow}c^\dagger_{I\uparrow}\ket{0}_I\\*
\ket{\uparrow\downarrow}_{IV}&=&d^\dagger_{IV\uparrow}d^\dagger_{IV\downarrow}\ket{0}_{IV}=-d^\dagger_{IV\downarrow}d^\dagger_{IV\uparrow}\ket{0}_{IV}
\end{eqnarray}
and, being consistent with the different Rindler regions operators anticommutation relations,
\begin{equation}\label{notation3}
\nonumber \ket{s}_I\ket{s'}_{IV}=c^\dagger_{Is}d^\dagger_{IVs'}\ket{0}_{I}\ket{0}_{IV}=-d^\dagger_{IVs'}c^\dagger_{Is}\ket{0}_{I}\ket{0}_{IV}
\end{equation}
\begin{equation}\label{notation4}
d^\dagger_{IVs'}\biket{s}{0}=-\biket{s}{s'}.
\end{equation}

Now, it is useful to note that \eqref{Bogoliubov} could be expressed as two-modes squeezing transformation for each $k$ \cite{Alicefalls,AlsingSchul}
\begin{equation}\left(\!\begin{array}{c}
a_{s,k}\\
b^\dagger_{k,s}
\end{array}\!\right)=S\left(\!\begin{array}{c}
c_{I,k,s}\\
d^\dagger_{IV,-k,-s}
\end{array}\!\right)S^\dagger\end{equation}
where
\begin{equation}\label{squeez}
S=\exp\left[r\left(c_{I,k,s}^\dagger\, d_{IV,-k,-s}e^{-i\phi}+c_{I,k,s}\, d_{IV,-k,-s}^\dagger e^{i\phi}\right)\right]
\end{equation}
So, analogously to \cite{Alicefalls,AlsingSchul}, it is reasonable to postulate that the Minkowski vacuum is a Rindler two-mode particles/antiparticles squeezed state with opposite spin and momentum states in I and IV. Contrarily to \cite{AlsingSchul}, considering that the modes have spin, occupation number for each $k$ is allowed to be 2, being higher occupation numbers forbidden by Pauli exclusion principle. In the literature the analysis is restricted only to one mode of the Minkowski field, but we can restrict our analysis to some sector of the Minkowski vacuum \eqref{vacua}, defining for the particles sector
\begin{equation}\label{revacua}
\ket{\tilde 0}=\bigotimes_{k_1,\dots,k_n}\ket{0_k}
\end{equation}
such that the particle sector of \eqref{vacua} can be written as \mbox{$\ket 0=\ket{\tilde 0}\otimes\bigotimes_{p\neq k_1,\dots,k_n}\ket{0_p}$.}

In this fashion we are considering a discrete number $n$ of different modes $k_1,\dots,k_n$, so Minkowski vacuum should be expressed as a squeezed state in Rindler coordinates which is an arbitrary superposition of spins and momenta. This will be useful to discuss what would happen if we relax the single-mode approximation carried out often in the literature and let our detectors have a small mode spread.
\begin{eqnarray}\label{vacuumCOMP}
\nonumber \ket{\tilde 0}&=&C^0\biket{0}{0}+\sum_{\substack{s_1\\k_1}}C^1_{s_1,k_1} \biket{\tilde 1}{\tilde 1}\\*
\nonumber&&+\sum_{\substack{s_1,s_2\\k_1,k_2}}C^2_{s_1,s_2,k_1,k_2}\xi_{s_1,s_2}^{k_1,k_2} \biket{\tilde 2}{\tilde 2}+\dots\\*
\nonumber &&+\!\!\!\!\!\sum_{\substack{s_1,\dots,s_n\\k_1,\dots,k_n}}\!\!\!\!C^n_{s_1,\dots,s_n,k_1,\dots,k_n}\xi_{s_1,\dots,s_n}^{k_1,\dots,k_n} \biket{\tilde n}{\tilde n}+\dots\\*
\nonumber&&+\!\!\!\!\!\!\!\sum_{\substack{s_1,\dots,s_{2n}\\k_1,\dots,k_{2n}}}\!\!\!\!C^{2n}_{s_1,\dots,s_{2n},k_1,\dots,k_{2n}}\xi_{s_1,\dots,s_{2n}}^{k_1,\dots,k_{2n}} \biket{\widetilde{2n}}{\widetilde{2n}}\\*
\end{eqnarray}
Where, here, the notation is
\begin{equation}\label{notationmod}
\biket{\tilde m}{\tilde m}\!=\!\biket{s_1,\!k_1;\dots;\!s_m,\!k_m}{-s_1,\!-k_1;\dots;\!-s_m,\!-k_m}
\end{equation}
with
\begin{equation}\label{notationmod2}
\ket{s_1,\!k_1;\dots;\!s_n,\!k_n}_I=c^\dagger_{I,k_n,s_n}\dots c^\dagger_{I,k_1,s_1}\ket{0}_I
\end{equation}
being
\begin{equation}\label{notationmod}
\sum_{\substack{s_1,\dots,s_m\\k_1,\dots,k_m}}\bra{\tilde m'}_{IV}\bra{\tilde m'}_I\biket{\tilde m}{\tilde m}\!=(m!)^2
\end{equation}
and the symbol $\xi$ is
\begin{equation}\label{gi}
\xi_{s_1,\dots,s_m}^{k_1,\dots,k_m}\equiv\left\{\begin{array}{l}
0 \; \text{If } s_i=s_j \text{ and } k_i=k_j \quad i\neq j\\[1mm]
1\; \text{Otherwise}
\end{array}\right.
\end{equation}
which imposes Pauli exclusion principle constraints on the state (quantum numbers of fermions cannot coincide).

Notice a pair of aspects of this notation for the multimode case. First, in the series in \eqref{vacuumCOMP} all the possible orders of the operators are implicitly written. Due to the anticommutation relations of the fermionic operators, terms with different orderings of the creation operators are related, i.e.
\begin{eqnarray}\label{ordering}
\nonumber\biket{s_1,k_1;s_2,k_2}{-s_1,-k_1;-s_2,-k_2}&=&\\*
\biket{s_2,k_2;s_1,k_1}{-s_2,-k_2;-s_1,-k_1}
\end{eqnarray}
So, without loss of generality, we could choose not to write all the possible orderings of the operators in \eqref{vacuumCOMP}. The difference between taking all the possible orderings of the operators into account and taking only one representant is a factor $m!$ in the constants $C^m$. From \eqref{ordering} we can also see that the coefficients $C^m$ are symmetric with respect to $s_i,k_i$ index permutations.

Second, as there are only $n$ different modes $(k_1,\dots,k_n)$, the last summation in equation \eqref{vacuumCOMP} has only $(2n)!$ terms due to \eqref{gi}. These terms are all the different permutations of the creation operators for pairs of opposed spins for each mode. There would be only one summand--instead of $(2n)!$--in the simplified notation where we do not write all the different permutations of the operators but only one representant. It means that, in this simplified notation, the series of terms with $n$ pairs has the same summands as the series with the vacuum state (i.e. only one). Actually, in this notation--i.e. if we count all the different order permutations as only one-- the series with $C^{2n}$ to $C^{n+1}$ has exactly the same number of summands as the series with $C^{0}$ to $C^{n-1}$.

To obtain restrictions on the values of the coefficients $C^m$ we demand that the Minkowski vacuum has to be annihilated by the particle annihilator, $a_{k_0,s_0}\ket0=0$. Translating this into Rindler coordinates we have
\begin{equation}\label{anhRindler}
\left[\cos r\,c_{I,k_0,s_0}-e^{i\phi}\sin r\,d^\dagger_{IV,-k_0,-s_0}\right]\ket0=0
\end{equation}
where the vacuum state should be expressed in Rindler coordinates using \eqref{vacuumCOMP}.

As the elements \eqref{notationmod} form an orthogonal set, from \eqref{anhRindler} we see that all the terms proportional to different elements of the set should be zero simultaneously, which gives the following conditions on the coefficients
\begin{itemize}
\item $C^1_{s,k}$ as a function of $C^0$\\[-9mm]
\end{itemize}
\begin{eqnarray}
\label{01} C^1_{\uparrow,k_0}\cos r-C^0e^{i\phi}\sin r&=&0\\*
C^1_{\downarrow,k_0}\cos r-C^0e^{i\phi}\sin r&=&0 \label{02}
\end{eqnarray}
since equations \eqref{01},\eqref{02} should be satisfied $\forall k_0$, we obtain that $C^1_{\uparrow,k}=C^1_{\downarrow,k}=\text{const.}$ since $C^0$ does not depend on $k$ or $s$. We will denote $C^1_{s,k}\equiv C^1$.
\begin{itemize}
\item $C^2_{s_1,s_2,k_1,k_2}$ as a function of $C^1$\\[-9mm]
\end{itemize}
\begin{eqnarray}
\label{03}  C^1e^{i\phi}\sin r- 2C^2_{ss',k,k_0}\cos r&=&0\\*
\label{04}  C^1e^{i\phi}\sin r- 2C^2_{ss',k_0,k}\cos r&=&0
\end{eqnarray}
since equations \eqref{03}, \eqref{04} should be satisfied $\forall k_0$, we obtain that $C^2_{s_1,s_2,k_1,k_2}=C^2$ where $C^2$ does not depend on spins or momenta since $C^1$ does not depend on $k$ or $s$, the only dependence of the coefficients \eqref{vacuumCOMP} with $k_i$ and $s_i$ is given by the Pauli exclusion principle, this dependence comes through function \eqref{gi}.

In fact it is very easy to show that all the coefficients are independent of $s_i$ and $k_i$ --apart from the Pauli exclusion principle constraint.-- Using the fact that $C^0$ does not depend on $s_i$ and $k_i$ and noticing that by applying the annihilator on the vacuum state and equalling it to zero, we will always obtain the linear relationship between $C^{n}$ and $C^{n-1}$ given below.
\begin{itemize}
\item $C^m$ as a function of $C^{m-1}$\\[-9mm]
\end{itemize}
\begin{eqnarray}
\label{05}  C^{m-1}e^{i\phi}\sin r- m\,C^m\cos r&=&0\\*
\label{06}  C^{m-1}e^{i\phi}\sin r- m\,C^m\cos r&=&0
\end{eqnarray}
We finally obtain that $C^m$ is a constant which can be expressed as a function of $C^0$ as
\begin{equation}\label{coeff2}
C^n=\frac{C^0}{m!} e^{im\phi}\tan^m r
\end{equation}
And then, vacuum state \eqref{vacuumCOMP} can be expressed as
\begin{eqnarray}\label{vacuumCOMP2}
\nonumber \ket{\tilde 0}&=&C^0\biket{0}{0}+C^1\sum_{\substack{s_1\\k_1}} \biket{\tilde 1}{\tilde 1}\\*
\nonumber&&+C^2\sum_{\substack{s_1,s_2\\k_1,k_2}}\xi_{s_1,s_2}^{k_1,k_2} \biket{\tilde 2}{\tilde 2}+\dots\\*
\nonumber &&+C^n\!\!\!\!\sum_{\substack{s_1,\dots,s_n\\k_1,\dots,k_n}}\!\!\!\!\xi_{s_1,\dots,s_n}^{k_1,\dots,k_n} \biket{\tilde n}{\tilde n}+\dots\\*
&&+C^{2n}\!\!\!\!\sum_{\substack{s_1,\dots,s_{2n}\\k_1,\dots,k_{2n}}}\!\!\!\!\xi_{s_1,\dots,s_{2n}}^{k_1,\dots,k_{2n}} \biket{\tilde n}{\tilde n}
\end{eqnarray}
where the only parameter not fixed yet is $C^0$.

We can now fix $C^0$ by imposing the normalization of the Minkowski vacuum in Rindler coordinates $\braket{0}{0}=1$, \cite{Alicefalls,AlsingSchul} this condition can be written as
\begin{equation}\label{series1}|C^0|^2\left[\sum_{m=0}^n\Upsilon_m\tan^{2m}r+\sum_{m=n+1}^{2n}\Upsilon_{2n-m}\tan^{2m}r\right]=1\end{equation}
Where
\begin{equation}\label{upsilon}\Upsilon_m=\sum_{\substack{s_1,\dots,s_m\\k_1,\dots,k_m}}\!\!\!\!\xi_{s_1,\dots,s_m}^{k_1,\dots,k_m}\end{equation}
and we have defined $\Upsilon_0\equiv 1$.
This expression gives (formally) the value of $C^0$ (except for a global phase) when considering the populated levels in an arbitrary number of modes of the field
\begin{equation}\label{series2}C^0=\left[\sum_{m=0}^n\Upsilon_m\tan^{2m}r+\sum_{m=n+1}^{2n}\Upsilon_{2n-m}\tan^{2m}r\right]^{-1/2}\end{equation}
We can see that if we take the limit $a\rightarrow0\Rightarrow r\rightarrow 0$ we recover the Minkowski Vacuum. We will come back to the behavior at the limits below.

Notice that the state \eqref{vacuumCOMP2} is only normalizable in this discrete limit. This comes about because the Minkowski and Rindler representations are not unitary equivalent (there is not unitary operator connecting the two vacua)\footnote{this is a old-known problem (see reference \cite{Takagi} chapter 2)}. It prevents us from taking the continuous limit in the expression \eqref{vacuumCOMP} but does not invalidate the treatment as equation \eqref{vacuumCOMP2} can be considered as a superposition of a finite number of individual modes which are perfectly well defined \cite{Takagi}.

To address the problem of showing how the presence of spin degrees of freedom affects the entanglement between accelerated observers, it is useful to use the single mode approximation \cite{Alsingtelep,AlsingMcmhMil} in the same way that in \cite{AlsingSchul,Alicefalls}. This is valid if we consider Rob's detector so sensitive to a single particle mode in region I that we can approximate the frequency $\omega_A=k_{0_A}$ observed by Alice to be the same frequency observed by Rob, $\omega_A\sim\omega_R$. As a consequence, the populated levels we are looking at are in this single frequency \cite{AlsingSchul} (See also discussion in \cite{AlsingMcmhMil}) so we can consider the sums over $k_i$ in \eqref{vacuumCOMP2} just like a sum of only one mode $k=\omega_A$. This is equivalent to restricting the analysis to the sector $\ket{0_k}$ of the complete vacuum \eqref{vacua}. Since the goal of this work is to show the effect of spin degrees of freedom on the entanglement for non-inertial observers, this approximation allows us to compare our results with previous literature on scalar and spinless fermion fields \cite{Alicefalls,AlsingSchul}.

We have to notice that since the observer Rob is accelerated, his possible measurements are affected by a Doppler-like effect. Given that the velocity of the observer is $\hat v = \hat a \hat t = \hat t/\hat z=\tanh(a t)$, the Doppler effect will shift the sensitivity peak of the detector. Namely, if at the instant $t=0$ his detector is sharply tuned to a frequency $\omega_D=\omega_A$, to compensate this Doppler effect at some instant $t=\tau$ the detector should be tuned to the frequency $\omega_D=e^{a\tau}\omega_A$. This implies that any detector will eventually become insensitive to the populated levels of the Minkowski field. In order to do the theoretical analysis in this work, we can consider either that his detector can be sharply tuned to the frequency $e^{a\tau}\omega_A$ for each instant, or that Rob has a set of individual detectors each one sharply tuned to the proper frequency for each instant.

Carrying out this single-mode approximation, the Minkowski vacuum for a single mode is a Rindler two-mode particles/antiparticles squeezed state with opposite spin states in I and IV. Considering that the modes have spin, occupation number is allowed to be 2 for each $k$, being higher occupation numbers forbidden by Pauli exclusion principle. As a consequence
\begin{eqnarray}\label{vacuum0}
\nonumber \ket{0}&=&V\biket{0}{0}+A\biket{\uparrow}{\downarrow}+B\biket{\downarrow}{\uparrow}\\*
&&+ C\biket{\pa}{\pa}.
\end{eqnarray}
Notice that $V$ is the analogous to $C^0$, $A$ and $B$ are analogous to $C^1_{\uparrow}$ and $C^1_{\downarrow}$ respectively and $C$ is analogous to $C^2$ in the expression \eqref{vacuumCOMP} but considering only one representant of all the 2 possible orders for the pair .

To obtain the values of the coefficients $V,A,B,C$ we demand that the Minkowski vacuum has to be annihilated by the particle annihilator, $a_s\ket0=0$. Translating this into Rindler coordinates we have
\begin{eqnarray}\label{annihil2}
\nonumber 0&=&\left[\cos r\,c_{I,s}-e^{i\phi}\sin r\,d^\dagger_{IV,-s}\right]\left[V\biket{0}{0}\right.\\*
&+&\left.A\biket{\uparrow}{\downarrow}+B\biket{\downarrow}{\uparrow}+ C\biket{\pa}{\pa}\right]
\end{eqnarray}
which implies
\begin{eqnarray}\label{annihil2}
\nonumber 0&=&\cos r\left[A\biket{0}{\downarrow}+B\delta_{s\downarrow}\biket{0}{\uparrow}\right.\\*
\nonumber &&+\left.C\left(\delta_{s\uparrow}\biket{\downarrow}{\pa}-\delta_{s\downarrow}\biket{\uparrow}{\pa}\right)\right]\\*
\nonumber&&-e^{i\phi}\sin r\left[V\biket{0}{-s}-A\delta_{s\downarrow}\biket{\uparrow}{\pa}\right.\\*
&&\left.+B\delta_{s\uparrow}\biket{\downarrow}{\pa}\right].
\end{eqnarray}

This equation gives 4 conditions (two for each value of $s$), although only 3 of them are independent
\begin{equation}
\left.\begin{array}{lcr}
A\cos r - Ve^{i\phi}\sin r&=&0\\
C\cos r - Be^{i\phi}\sin r&=&0\\
B\cos r - Ve^{i\phi}\sin r&=&0\\
C\cos r - Ae^{i\phi}\sin r&=&0\\
\end{array}\right\}\Rightarrow \begin{array}{ll}A=B=V e^{i\phi}\tan r\\
C=V e^{2i\phi}\tan^2 r
\end{array}
\end{equation}
To fix $V$ we impose the normalization relation for each field mode $\braket{0}{0}=1\Rightarrow |V|^2=1-|A|^2-|B|^2-|C|^2$, imposing this we finally obtain the values of the vacuum coefficients.
\begin{equation}\label{vaccoef}
\begin{array}{lcl}
V&=&\cos^2 r\\
A&=&e^{i\phi}\sin r\,\cos r \\
B&=&e^{i\phi}\sin r\,\cos r\\
C&=&e^{2i\phi}\sin^2 r\\
\end{array}
\end{equation}
Notice that comparing this result with expressions \eqref{coeff2} and \eqref{series2}, as we have truncated the series in \eqref{series2}, the value of $V$ will be different from the case when more than one mode is considered --$C^0$ instead of $V$--. If we restrict the series on $m$ to only one mode $n=1$ in \eqref{series1}, we obtain that $C^0\rightarrow1/(1+\tan^2 r)=\cos^2r$ and we get then proper values for $A=B=C^1$ and $C=2! C^2$.

Since from \eqref{defr} $a\rightarrow\infty\Rightarrow r\rightarrow \pi/4$, comparing $V$ with \eqref{series2}, we can see that, while under the single-mode approximation the limit of infinite acceleration leads to a finite distribution of the Minkowski vacuum over Rindler states, when considering the multimode Rindler expression for the vacuum state \eqref{vacuumCOMP2} the combined limit of $n\rightarrow\infty$ and $a\rightarrow\infty\Rightarrow r\rightarrow\pi/4$ leads to a complete fading away of the amplitudes over all the Rindler modes as $C_0\rightarrow 0$, which may not be the case for finite $a$. This is beyond the scope of this article but we will discuss in the conclusion that it may have very strong implications on the entanglement of fermionic fields for accelerated observers.

So finally, under the single mode approximation, the Minkowski vacuum state in Rindler coordinates is as follows
\begin{eqnarray}\label{vacuum}
\nonumber \ket{0}&=&\cos^2 r\,\biket{0}{0}+e^{i\phi}\sin r\,\cos r\left(\biket{\uparrow}{\downarrow}\right.\\*
&&\left.+\biket{\downarrow}{\uparrow}\right)+e^{2i\phi}\sin^2 r\,\biket{\pa}{\pa}
\end{eqnarray}

Now we have to build the one particle (of spin $s$) state in Rindler coordinates. It can be readily done by applying the Minkowski particle creation operator to the vacuum state $\ket{s}=a^\dagger_s\ket0$, and translating it into Rindler coordinates:
\begin{eqnarray}\label{onepart1}
\nonumber \ket{s}&=&\left[\cos r\,c^\dagger_{I,s}-e^{-i\phi}\sin r\,d_{IV,-s}\right]\left[\cos^2 r\biket{0}{0}\right.\\*
\nonumber&&+e^{i\phi}\sin r\,\cos r\left(\biket{\uparrow}{\downarrow}+\biket{\downarrow}{\uparrow}\right)\\*
&&\left.+e^{2i\phi}\sin^2 r\,\biket{\pa}{\pa}\right]
\end{eqnarray}
That means
\begin{eqnarray}\label{onepart2}
\nonumber\ket\uparrow&=&\cos r \biket{\uparrow}{0}+e^{i\phi}\sin r\biket{\pa}{\uparrow}\\*
\ket\downarrow&=&\cos r \biket{\downarrow}{0}-e^{i\phi}\sin r\biket{\pa}{\downarrow}
\end{eqnarray}

The three Minkowski states $\ket0,\ket\uparrow,\ket\downarrow$ correspond to the particle field of mode $k$ observed by Alice. However, since Rob is experiencing a uniform acceleration he will not be able to access to field modes in the causally disconnected region IV, hence, Rob must trace over that inaccessible region as it is unobservable.

Specifically, when Rob is in region I of Rindler space-time and Alice observes the vacuum state, Rob could only observe a non-pure partial state given by $\rho_R=\tr_{IV}\left(\ket{0}\bra{0}\right)$ that is
\begin{eqnarray}\label{partialvacuum}
\nonumber\rho_R&=&\cos^4 r\ket{0}_I\!\!\bra{0}+\sin^2 r\,\cos^2 r\left(\ket{\uparrow}_I\!\!\bra{\uparrow}\right.\\*
&&\left.+\ket{\downarrow}_I\!\!\bra{\downarrow}\right)+\sin^4 r \ket{\pa}_I\!\bra{\pa}
\end{eqnarray}
But while Alice would observe the vacuum state of mode $k$, Rob would observe certain statistical distribution of particles. The expected value of Rob's number operator on the Minkowski vacuum state is given by
\begin{eqnarray}
\nonumber\langle N_R\rangle&=&\ematriz{0}{N_R}{0}=\tr_{I,IV}\left(N_R\proj{0}{0}\right)=\tr_{I}\left(N_R\rho_R\right)\\*
&=&\tr_{I}\left[\left(c_{I\uparrow}^\dagger c_{I\uparrow}+c_{I\downarrow}^\dagger c_{I\downarrow}\right)\rho_R\right]
\end{eqnarray}
Substituting the expression \eqref{partialvacuum} we obtain
\begin{equation}
\langle N_R\rangle=2\sin^2 r
\end{equation}
using \eqref{defr} we obtain that
\begin{equation}\label{Unruh}
\langle N \rangle=2\frac{1}{e^{2\pi\omega c/a}+1}=2\frac{1}{e^{\hslash\omega/K_BT}+1}
\end{equation}
where $k_B$ is the Boltzmann's constant and
\begin{equation}
T=\frac{\hslash\, a}{2\pi k_B c}
\end{equation}
is the Unruh temperature.

Equation \eqref{Unruh} is known as the Unruh effect \cite{DaviesUnr,Unruh}, which shows that, for a two-dimensional space-time, an uniformly accelerated observer in region I detects a thermal Fermi-Dirac distribution when he observes the Minkowski vacuum. We obtain a factor 2 contrarily to Ref. \cite{AlsingSchul} due to the degeneracy factor $2S+1$.

\section{Spin entanglement with an accelerated partner}\label{sec5}

In previous works \cite{Alicefalls,AlsingSchul} it was studied how Unruh decoherence affects occupation number entanglement in bipartite states as
\begin{equation}\label{alsingst}
\ket{\Psi}=\frac{1}{\sqrt2}(\ket{00}+\ket{11})
\end{equation}where the figures inside the kets represent occupation number of Alice and Rob modes respectively, barring any reference to the spin of the field modes.

Here, where we have included the spin structure of each mode in our setting from the very beginning, it is possible to study the effects of acceleration in spin entanglement decoherence, which is different from the mere occupation number entanglement.

First of all, we build a general bipartite state that could be somehow analogous to state \eqref{alsingst} studied in \cite{AlsingSchul}, limiting the occupation number to 1 but including the spins of each mode.
\begin{eqnarray}\label{genstate}
\nonumber\ket{\Psi}&=&\mu\ket{0_A}\ket{0_R}+\alpha\ket{\uparrow_A}\ket{\uparrow_R}+\beta\ket{\uparrow_A}\ket{\downarrow_R}\\*
&&+ \gamma\ket{\downarrow_A}\ket{\uparrow_R}+\delta\ket{\downarrow_A}\ket{\downarrow_{R}}
\end{eqnarray}
with $\mu=\sqrt{1-|\alpha|^2-|\beta|^2-|\gamma|^2-|\delta|^2}$. The subscripts $A,R$ indicate the modes associated with Alice and Rob respectively. We will suppress the labels $A,R$  from now on, and we will understand that the first character in a ket or a bra corresponds to Alice and the second to Rob: $\ket{s,s'}=\ket{s_A}\ket{s'_R}$

This general setting \eqref{genstate} allows us to study in this section what happens with spin entanglement under acceleration of Rob and also what happens with the occupation number entanglement when considering sates analogous to \eqref{alsingst} but taking the spin structure into account. It will also allow us to discuss, in section \ref{sec6}, the implications of tracing over spins and study only the entanglement on the occupation number degree of freedom compared with \cite{AlsingSchul}.

The density matrix in Minkowski coordinates for the state \eqref{genstate} is
\begin{eqnarray}\label{Minkowdens}
\nonumber\rho^M&=&\mu^2\proj{0,0}{0,0}+\mu\alpha^*\proj{0,0}{\uparrow,\uparrow}+\mu\beta^*\proj{0,0}{\uparrow,\downarrow}\\*
\nonumber &&+\mu\gamma^*\proj{0,0}{\downarrow,\uparrow}+\mu\delta^*\proj{0,0}{\downarrow,\downarrow}+|\alpha|^2\proj{\uparrow,\uparrow}{\uparrow,\uparrow}\\*
\nonumber &&+\alpha\beta^*\proj{\uparrow,\uparrow}{\uparrow,\downarrow}+\alpha\gamma^*\proj{\uparrow,\uparrow}{\downarrow,\uparrow}+\alpha\delta^*\proj{\uparrow,\uparrow}{\downarrow,\downarrow}\\*
\nonumber&&+|\beta|^2\proj{\uparrow,\downarrow}{\uparrow,\downarrow}+\beta\gamma^*\proj{\uparrow,\downarrow}{\downarrow,\uparrow}+\beta\delta^*\proj{\uparrow,\downarrow}{\downarrow,\downarrow}\\*
\nonumber&&+|\gamma|^2\proj{\downarrow,\uparrow}{\downarrow,\uparrow}+\gamma\delta^*\proj{\downarrow,\uparrow}{\downarrow,\downarrow}+|\delta|^2\proj{\downarrow,\downarrow}{\downarrow,\downarrow}\\*
&&+ \text{n.d.H.c.}
\end{eqnarray}
where n.d.H.c means non-diagonal Hermitian conjugate, and represents the Hermitian conjugate only for the non-diagonal elements.

Computing the density matrix, taking into account that Rob is constrained to region I of Rindler space-time, requires to rewrite Rob's mode in terms of Rindler modes and to trace over the unobservable Rindler's region IV.

In appendix \ref{appen} we compute each term of \eqref{Minkowdens} in Rindler's coordinates and trace over the unobserved region IV. Using \eqref{trazapa1}, \eqref{trazapa2}, \eqref{trazapa3} we can easily compute the density matrix for Alice and Rob from \eqref{Minkowdens} since $\rho_{AR}=\tr_{IV}\rho_M$, resulting in the long expression
\begin{eqnarray}\label{generaldensmat}
\nonumber\rho_{AR}&=&\mu^2\Big[\cos^4r\proj{0,0}{0,0}+\sin^2r\cos^2r\left(\proj{0,\uparrow}{0,\uparrow}\right.\\*
\nonumber&&\left.+\proj{0,\downarrow}{0,\downarrow}\right)+\sin^4r\proj{0,\pa}{0,\pa}\Big]+\mu\cos^3r\\*
\nonumber&&\times\Big[\alpha^*\proj{0,0}{\uparrow,\uparrow}+\beta^*\proj{0,0}{\uparrow,\downarrow}+\gamma^*\proj{0,0}{\downarrow,\uparrow}\\*
\nonumber&&+\delta^*\proj{0,0}{\downarrow,\downarrow}\Big]+\mu\sin^2r\,\cos r\Big[\alpha^*\proj{0,\downarrow}{\uparrow,\pa}\\*
\nonumber&&-\beta^*\!\proj{0,\uparrow}{\uparrow,\pa}\!+\!\gamma^*\!\proj{0,\downarrow}{\downarrow,\pa}\!-\!\delta^*\!\proj{0,\uparrow}{\downarrow,\pa}\Big]\\*
\nonumber&&+\cos^2 r\Big[|\alpha|^2\proj{\uparrow,\uparrow}{\uparrow,\uparrow}+\alpha\beta^*\proj{\uparrow,\uparrow}{\uparrow,\downarrow}+\alpha\gamma^*\\*
\nonumber&&\times\proj{\uparrow,\uparrow}{\downarrow,\uparrow}+\alpha\delta^*\proj{\uparrow,\uparrow}{\downarrow,\downarrow}+|\beta|^2\proj{\uparrow,\downarrow}{\uparrow,\downarrow}\\*
\nonumber&&+\beta\gamma^*\proj{\uparrow,\downarrow}{\downarrow,\uparrow}+\beta\delta^*\proj{\uparrow,\downarrow}{\downarrow,\downarrow}+|\gamma|^2\proj{\downarrow,\uparrow}{\downarrow,\uparrow}\\*
\nonumber&&+\gamma\delta^*\proj{\downarrow,\uparrow}{\downarrow,\downarrow}+|\delta|^2\proj{\downarrow,\downarrow}{\downarrow,\downarrow}\Big]+\sin^2r\\*
\nonumber&&\times\Big[\left(|\alpha|^2+|\beta|^2\right)\proj{\uparrow,\pa}{\uparrow,\pa}+\left(|\gamma|^2+|\delta|^2\right)\\*
\nonumber&&\times\proj{\downarrow,\pa}{\downarrow,\pa}+\left(\alpha\gamma^*+\beta\delta^*\right)\proj{\uparrow,\pa}{\downarrow,\pa}\Big]\\*
&&+\text{n.d.H.c.}
\end{eqnarray}
Here the notation is the same than in the r.h.s. of \eqref{trazapa1}: $\proj{a,r}{a',r'}=\ket{a_A}\ket{r_R}_I\bra{a'_A}\,{_{I}\!\!\bra{r'_R}}$. Notice that the state, which in Minkowski coordinates is pure, gets mixed when the observer Rob is accelerated.

Equation \eqref{generaldensmat} will be our starting point, from which we will study different entanglement settings and how Unruh decoherence affects them.

To begin with we will compute how acceleration affects the entanglement of spin Bell states when Alice and Rob share a maximally entangled spin state and Rob accelerates. In Minkowski coordinates that means choosing specific coefficients in \eqref{genstate}, particularly, for Bell states we should choose
\begin{eqnarray}
\ket{\phi^\pm}\Rightarrow \alpha=\pm\delta=\frac{1}{\sqrt{2}}\\*
\ket{\psi^\pm}\Rightarrow \beta=\pm\gamma=\frac{1}{\sqrt{2}}
\end{eqnarray}
and the rest of the other coefficients equal to zero. For such states in Minkowski coordinates, the density matrix of Alice and Rob considering that Rob undergoes an acceleration $a$ is obtained from \eqref{generaldensmat}:
\begin{eqnarray}\label{Phibell}
\nonumber\rho^{\phi^\pm}_{AR}&=&\frac12\Big[\cos^2 r\Big(\proj{\uparrow,\uparrow}{\uparrow,\uparrow}\pm\proj{\uparrow,\uparrow}{\downarrow,\downarrow}\pm\proj{\downarrow,\downarrow}{\uparrow,\uparrow}\\*
\nonumber&&+\proj{\downarrow,\downarrow}{\downarrow,\downarrow}\Big)+\sin^2 r\Big(\proj{\uparrow,\pa}{\uparrow,\pa}\\*
&&+\proj{\downarrow,\pa}{\downarrow,\pa}\Big)\Big]
\end{eqnarray}
\begin{eqnarray}\label{Psibell}
\nonumber\rho^{\psi^\pm}_{AR}&=&\frac12\Big[\cos^2 r\Big(\proj{\uparrow,\downarrow}{\uparrow,\downarrow}\pm\proj{\uparrow,\downarrow}{\downarrow,\uparrow}\pm\proj{\downarrow,\uparrow}{\uparrow,\downarrow}\\*
\nonumber&&+\proj{\downarrow,\uparrow}{\downarrow,\uparrow}\Big)+\sin^2 r\Big(\proj{\uparrow,\pa}{\uparrow,\pa}\\*
&&+\proj{\downarrow,\pa}{\downarrow,\pa}\Big)\Big]
\end{eqnarray}

Notice that, in this case, Alice would have a qubit and Rob would have a qutrit, since for his mode he could have three different possible orthogonal states: particle spin-up, particle spin-down and particle pair.

To characterize its entanglement we will use the negativity \cite{Negat} normalized to one (we can multiply it by a constant in order to have negativity equal to one for a maximally entangled state), Therefore, to have negativity equal to 1 for a Bell state we define it as twice the addition of all the negative eigenvalues of the partial transpose density matrix --which consists on transposing Rob's qutrits--
\begin{eqnarray}
\nonumber\rho^{\phi^\pm pT}_{AR}\!\!&=&\frac12\Big[\cos^2 r\Big(\proj{\uparrow,\uparrow}{\uparrow,\uparrow}\pm\proj{\uparrow,\downarrow}{\downarrow,\uparrow}\pm\proj{\downarrow,\uparrow}{\downarrow,\uparrow}\\*
\nonumber&&+\proj{\downarrow,\downarrow}{\downarrow,\downarrow}\Big)+\sin^2 r\Big(\proj{\uparrow,\pa}{\uparrow,\pa}\\*
&&+\proj{\downarrow,\pa}{\downarrow,\pa}\Big)\Big]
\end{eqnarray}
\begin{eqnarray}
\nonumber\rho^{\psi^\pm pT}_{AR}\!\!&=&\frac12\Big[\cos^2 r\Big(\proj{\uparrow,\downarrow}{\uparrow,\downarrow}\pm\proj{\uparrow,\uparrow}{\downarrow,\downarrow}\pm\proj{\downarrow,\downarrow}{\uparrow,\uparrow}\\*
\nonumber&&+\proj{\downarrow,\uparrow}{\downarrow,\uparrow}\Big)+\sin^2 r\Big(\proj{\uparrow,\pa}{\uparrow,\pa}\\*
&&+\proj{\downarrow,\pa}{\downarrow,\pa}\Big)\Big]
\end{eqnarray}
We can write $\rho^{\phi^\pm pT}_{AR}$ matricially in the basis $\left\{\ket{\uparrow,\uparrow},\ket{\uparrow,\downarrow},\ket{\downarrow,\uparrow},\ket{\downarrow,\downarrow},\ket{\uparrow,\pa},\ket{\downarrow,\pa}\right\}$
\begin{equation}
\frac12\left(\!\begin{array}{cccccc}
\cos^2 r & 0 &0 &0 & 0 & 0\\
0&0 &\pm\cos^2 r &0 & 0 & 0 \\
0  &\pm\cos^2 r  &0 &0 &0 &0 \\
0  & 0 &0 &\cos^2 r &0 &0 \\
0 & 0 &0&0& \sin^2 r &0 \\
0& 0& 0&0&0& \sin^2 r \\
\end{array}\!\right)
\end{equation}
which have the same expression than $\rho^{\psi^\pm pT}_{AR}$ in the basis $\left\{\ket{\uparrow,\downarrow},\ket{\uparrow,\uparrow},\ket{\downarrow,\downarrow},\ket{\downarrow,\uparrow},\ket{\uparrow,\pa},\ket{\downarrow,\pa}\right\}$. Therefore the four Bell states will have the same eigenvalues which are:
\begin{eqnarray}
\nonumber\lambda_1=\lambda_2=\lambda_3=\frac12\cos^2 r\\*
\lambda_4=\lambda_5=\frac12\sin^2 r\\*
\nonumber\lambda_6=-\frac12\cos^2r
\end{eqnarray}

Since $r=\arctan\left(e^{-\pi\frac{\omega c}{a}}\right)$ $a\rightarrow0\Rightarrow r\rightarrow 0$ and $a\rightarrow\infty\Rightarrow r\rightarrow \pi/4$ so that $\lambda_6$ is negative for all values of the acceleration. This implies, using Peres criterion \cite{PeresCriterion}, that the spin Bell states will be always entangled even in the limit of infinite acceleration.

We can readily evaluate the entanglement at the limits $a\rightarrow0$ and $a\rightarrow\infty$ if we compute the negativity (normalized to one for maximally entangled states), that is to say
\begin{equation}\label{negativity}
\mathcal{N}=2\sum_{\lambda_i<0}|\lambda_i|
\end{equation}
Applied to our states we obtain that
\begin{equation}
\mathcal{N}(r)=\cos^2 r
\end{equation}
In the limit $a\rightarrow0$ we obtain $\mathcal{N}=1$ which is an expected result since $a\rightarrow0$ is the inertial limit.

However, in the limit $a\rightarrow\infty$ we obtain $\mathcal{N}=\frac{1}{2}$, which implies that spin entanglement degrades due to the Unruh effect. Fig. 2 shows the negativity as a function of the acceleration of Rob.
\begin{figure}\label{fig2}
\includegraphics[width=.45\textwidth]{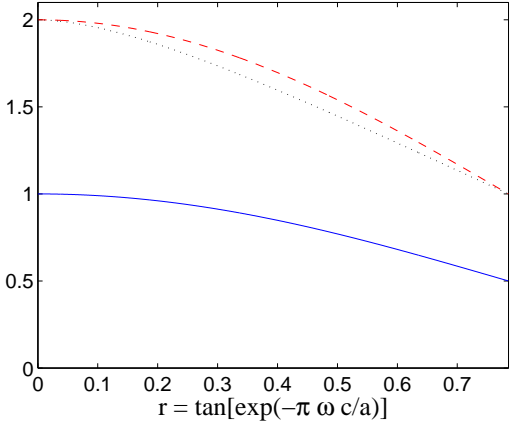}
\caption{Negativity and mutual information as a function of the acceleration of Rob when $R$ and $A$ share a maximally entangled state in Minkowski coordinates. Red dashed line is mutual information for all the spin Bell states. Black dotted line is mutual information for the Minkowski state $\frac{1}{\sqrt2}\left(\ket{00}+\ket{\uparrow\downarrow}\right)$ and blue solid line is negativity for both, Bell spin states and $\frac{1}{\sqrt2}\left(\ket{00}+\ket{\uparrow\downarrow}\right)$}
\end{figure}

The mutual information, which takes into account quantum and classical correlations, is
\begin{equation}\label{Imutua}
I_{AR}=S_{A}+S_{R}-S_{AR}
\end{equation}
where $S_{A,R}$ are the Von Neumann entropies of the partial state of Alice and Rob and $S_{AR}$ is the entropy of the whole state.

For \eqref{Phibell} and \eqref{Psibell}, the partial states of Alice and Rob ($\rho_A=\tr_R\rho_{AR}$, $\rho_R=\tr_A\rho_{AR}$) can be expressed matricially as
\begin{equation}
\rho_A=\frac12\left(\!\begin{array}{cc}
1 & 0 \\
0 & 1
\end{array}\!\right)
\end{equation}
in the basis $\{\ket{\uparrow},\ket{\downarrow}\}$
\begin{equation}
\rho_R=\frac12\left(\!\begin{array}{ccc}
\cos^2 r & 0 \\
0 & \cos^2 r &0\\
0 &0 & 2\sin^2 r
\end{array}\!\right)
\end{equation}
in the basis $\{\ket{\uparrow},\ket{\downarrow},\ket{\pa}\}$
for all the Bell states, and
\begin{equation}
\rho_{AR}=\frac12\left(\!\begin{array}{cccc}
\cos^2 r & \pm\cos^2 r &0 & 0 \\
\pm\cos^2 r & \cos^2 r  &0 & 0 \\
0 &0 & \sin^2 r & 0 \\
0 &0 & 0& \sin^2 r  \\
\end{array}\!\right)
\end{equation}
for $\phi^{\pm}$ in the basis $\{\ket{\uparrow,\uparrow},\ket{\downarrow,\downarrow},\ket{\uparrow,\pa},\ket{\downarrow,\pa}\}$  and the same expression for $\psi^{\pm}$ in the basis $\{\ket{\uparrow,\downarrow},\ket{\downarrow,\uparrow},\ket{\uparrow,\pa},\ket{\downarrow,\pa}\}$. The entropies of these states are
\begin{eqnarray}\label{Entrop}
\nonumber S_A&=&1\\*
\nonumber S_R&=&-\cos^2r\,\log_2\left(\frac12\cos^2r\right)-\sin^2 r\,\log_2\left(\sin^2r\right)\\*
\nonumber S_{AR}&=&\cos^2r\,\log_2\left(\cos^2r\right)-\sin^2r\,\log_2\left(\frac12\sin^2r\right)\\*
\end{eqnarray}
and the mutual information is
\begin{equation}\label{mutual2}
I_{AR}=2\cos^2r
\end{equation}
Again we see that in the limit $a\rightarrow 0$ mutual information goes to 2 and in the limit of infinite acceleration it goes to 1. The behavior of the mutual information as a function of $a$ is shown in fig. 1.

In \cite{AlsingSchul} it is discussed that Pauli exclusion principle protects the on occupation number entanglement from decoherence, and some degree of entanglement is preserved even at the limit $a\rightarrow\infty$. Here we have obtained a similar result for the spin Bell states, showing that spin entanglement is also degraded by Unruh effect.

Next, we will study the case in which Alice and Rob share a different class of maximally entangled state. We consider that in Minkowski coordinates we have
\begin{equation}\label{minkstate2}
\ket{\Psi}=\frac{1}{\sqrt2}\left(\ket{0_A}\ket{0_R}+\ket{\uparrow_A}\ket{\downarrow_R}\right)
\end{equation}
which is a maximally entangled state that includes occupation number entanglement along with spin. We study this kind of states as a first analog to the state considered in previous literature \eqref{alsingst}. This state corresponds to the choice
\begin{eqnarray}\label{coef22}
\beta&=&\mu=\frac{1}{\sqrt2}\\*
\alpha&=&\gamma=\delta=0
\end{eqnarray}
in equation \eqref{generaldensmat}. The density matrix of such a state is
\begin{eqnarray}\label{minkstate2R}
\nonumber\rho&=&\frac12\Big[\cos^4r\proj{0,0}{0,0}+\sin^2r\cos^2r\left(\proj{0,\uparrow}{0,\uparrow}\right.\\*
\nonumber&&\left.+\proj{0,\downarrow}{0,\downarrow}\right)+\sin^4r\proj{0,\pa}{0,\pa}+\cos^3r\\*
\nonumber&&\times\left(\proj{0,0}{\uparrow,\downarrow}+\proj{\uparrow,\downarrow}{0,0}\right)-\sin^2r\,\cos r\\*
\nonumber&&\times\left(\proj{0,\uparrow}{\uparrow,\pa}+\proj{\uparrow,\pa}{0,\uparrow}\right)+\cos^2r\proj{\uparrow,\downarrow}{\uparrow,\downarrow}\\*
&&+\sin^2r\proj{\uparrow,\pa}{\uparrow,\pa}\Big]
\end{eqnarray}
Notice the significant difference from the Bell spin states; considering that Rob accelerates means that, this time, Alice has a qubit and Rob has a qu4it. Hence, negativity acts only as a measure of distillable entanglement, and does not account for the possible bound entanglement the system would have \cite{HorodeckiBound}. Since in Rindler coordinates the state \eqref{minkstate2R} is qualitatively different from the Minkowski Bell states \eqref{Phibell}, \eqref{Psibell}, it is therefore worthwhile to study its entanglement and the mutual information degradation as Rob accelerates.

The partial transpose of \eqref{minkstate2R} $\sigma=\rho^{pT}$ is
\begin{eqnarray}\label{minkstate2R2}
\nonumber\sigma&=&\frac12\Big[\cos^4r\proj{0,0}{0,0}+\sin^2r\cos^2r\left(\proj{0,\uparrow}{0,\uparrow}\right.\\*
\nonumber&&\left.+\proj{0,\downarrow}{0,\downarrow}\right)+\sin^4r\proj{0,\pa}{0,\pa}+\cos^3r\\*
\nonumber&&\times\left(\proj{0,\downarrow}{\uparrow,0}+\proj{\uparrow,0}{0,\downarrow}\right)-\sin^2r\,\cos r\\*
\nonumber&&\times\left(\proj{0,\pa}{\uparrow,\uparrow}+\proj{\uparrow,\uparrow}{0,\pa}\right)+\cos^2r\proj{\uparrow,\downarrow}{\uparrow,\downarrow}\\*
&&+\sin^2r\proj{\uparrow,\pa}{\uparrow,\pa}\Big]
\end{eqnarray}
which is an $8\times8$ matrix. $\sigma$ is diagonal by blocks with eigenvalues
\begin{eqnarray}\label{eig2}
\nonumber\lambda_1&=&\frac12\cos^4r\\*
\nonumber\lambda_2&=&\frac12\cos^2r\sin^2r\\*
\lambda_3&=&\frac12 \sin^2 r\\*
\nonumber\lambda_4&=&\frac12\cos^2r\\*
\nonumber\lambda_{5,6}&=&\frac{1}{4}\left(\sin^2r\cos^2r\pm\sqrt{\sin^4r\cos^4r+4\cos^6r}\right)\\*
\nonumber\lambda_{7,8}&=&\frac14\left(\sin^4r\pm\sqrt{\sin^8r+4\sin^4r\,\cos^2r}\right)
\end{eqnarray}
As we can see, $\lambda_8$ is non-positive and $\lambda_6$ is negative for all values of $a$, therefore the state will always preserve some degree of distillable entanglement. If we calculate the negativity we will obtain
\begin{equation}
\mathcal{N}(r)=\cos^2r
\end{equation}
which means that for this case, distillable entanglement behaves equally than in the previous case, and negativity on fig. 1 is equally valid for this state.

Finally we compute the mutual information of the system whose partial matrices are expressed as
\begin{equation}
\rho_A=\frac12\left(\!\begin{array}{cc}
1 & 0 \\
0 & 1
\end{array}\!\right)
\end{equation}
in the basis $\{\ket{0},\ket{\uparrow}\}$. $\rho_R=$
\begin{equation}
\frac12\left(\!\begin{array}{cccc}
\cos^4 r & 0 & 0 &0\\
0 & \!\!\!\sin^2r\,\cos^2 r &0 & 0\\
0 &0 & \!\!\!\!\cos^2r\left(\sin^2 r+1\right) & 0\\
0 & 0& 0& \!\!\!\!\!\sin^2r\left(\sin^2r+1\right)
\end{array}\!\right)
\end{equation}
in the basis $\{\ket{0},\ket{\uparrow},\ket{\downarrow},\ket{\pa}\}$. The eigenvalues of the whole system $6\times6$ matrix $\rho_{AR}$ are
\begin{eqnarray}
\nonumber \lambda_1&=&\lambda_2=0\\*
\nonumber \lambda_3&=&\frac12\sin^2r\cos^2r\\*
\nonumber \lambda_4&=&\frac12\sin^4r\\*
\nonumber \lambda_5&=&\frac12\cos^2r\left(1+\cos^2r\right)\\*
\lambda_6&=&\frac12\sin^2r\left(1+\cos^2r\right)
\end{eqnarray}
In this case the mutual information as a function of $a$ is not proportional to the negativity. Hence, it is different from the Bell states \eqref{Phibell},\eqref{Psibell}. As it can be seen in Fig. 1 the value of mutual information for \eqref{Phibell}, \eqref{Psibell} and \eqref{minkstate2R} coincide at the limits $a\rightarrow0,a\rightarrow\infty$, but are different in between, obtaining that $I_{AR}^{\text{SpinBell}}\ge I_{AR}^{\text{ModeBell}}$.

To conclude this section we stress that the same results will be obtained if the state $\ket{\uparrow,\downarrow}$ in \eqref{minkstate2} is replaced by any other 1 particle bipartite spin state $\ket{s,s'}$.

\section{Occupation number entanglement with an accelerated partner and spin $1/2$ fermions}\label{sec6}

The previous work \cite{AlsingSchul} on occupation number entanglement between accelerated partners ignored the spin structure of the Dirac field modes. It is not possible to straightforwardly translate a state like \eqref{genstate} into mere occupation number states. This comes about because for a state like \eqref{genstate} the bipartite vacuum component does not have individual spin degrees of freedom as the other components do. In other words, by including the vacuum state in the superposition \eqref{genstate}, the Hilbert space ceases to be factorable in terms of individual spin times particle occupation number subspaces.

On the other hand, the bipartite vacuum is a well defined total spin singlet. Hence, the Hilbert space is factorable with respect to the total spin of the system $A-R$ and the occupation number subspaces. Accordingly, to reduce the spin information in the general density matrix \eqref{generaldensmat} we will be forced to consider a factorization of the Hilbert space as the product of the total spin and occupation number subspaces.

If we do such a factorization we could consider that we are not able to access to the information of the total spin of the system $A-R$ and then, we should trace over total spin degree of freedom.

The equivalence between the standard basis (occupation number-individual spin) and the new basis (occupation number-total spin) is given\footnote{The pair state in the same mode can only be a singlet of total spin due to anticommutation relations of fermionic fields} in equations \eqref{e1} and \eqref{e2}
\begin{equation}\label{e1}
\begin{array}{cc}
\ket{0,0}=\ket{00}\ket{S}& \ket{0,\pa}=\ket{02}\ket{S}\\*[1.5mm]
\ket{0,\uparrow}=\ket{01}\ket{D_+}& \ket{0,\downarrow}=\ket{01}\ket{D_-}\\*[1.5mm]
\ket{\uparrow,0}=\ket{10}\ket{D_+}&\ket{\downarrow,0}=\ket{10}\ket{D_-}\\*[1.5mm]
\ket{\uparrow,\pa}=\ket{12}\ket{D_+}&\ket{\downarrow,\pa}=\ket{12}\ket{D_-}\\*[1.5mm]
\ket{\uparrow,\uparrow}=\ket{11}\ket{T_+}&\ket{\downarrow,\downarrow}=\ket{11}\ket{T_-}\\*[1.5mm]
\end{array}
\end{equation}
\begin{equation}\label{e2}
\begin{array}{c}
\ket{\uparrow,\downarrow}=\frac{1}{\sqrt{2}}\ket{11}\left[\ket{T_0}+\ket{S}\right]\\*[1.5mm]
\ket{\downarrow,\uparrow}=\frac{1}{\sqrt{2}}\ket{11}\left[\ket{T_0}-\ket{S}\right]
\end{array}
\end{equation}
where we are using the basis $\ket{n_a\,n_b}\ket{J,J_z}$ and the triplets, doublets and the singlet are denoted as
\begin{eqnarray}
\nonumber\ket{T_+}&=&\ket{J=1,J_z=1}\\*
\nonumber\ket{T_-}&=&\ket{J=1,J_z=-1}\\*
\nonumber\ket{T_0}&=&\ket{J=1,J_z=0}\\*
\nonumber\ket{D_+}&=&\ket{J=1/2,J_z=1/2}\\*
\nonumber\ket{D_-}&=&\ket{J=1/2,J_z=-1/2}\\*
\nonumber\ket{S}&=&\ket{J=0,J_z=0}\\*
\end{eqnarray}
If we rewrite the general state \eqref{genstate} in this basis we obtain
\begin{eqnarray}\label{genstateb2}
\nonumber\ket{\Psi}&=&\mu\ket{00}\ket{S}+\alpha\ket{11}\ket{T_+}+\frac{\beta+\gamma}{\sqrt2}\ket{11}\ket{T_0}\\*
&&+ \frac{\beta-\gamma}{\sqrt2}\ket{11}\ket{S}+\delta\ket{11}\ket{T_-}
\end{eqnarray}
And the general state when Rob is accelerated \eqref{generaldensmat} in terms of this new basis after reducing the information on the total spin by tracing over this degree of freedom is
\begin{equation}
\rho^n_{AR}=\sum_{J,J_z}\bra{J,J_z}\rho_{AR}\ket{J,J_z}
\end{equation}
Which results in a state in the occupation number basis whose entanglement decoherence could be studied and compared with the results in reference \cite{AlsingSchul} in which spin is ignored:
\begin{eqnarray}\label{densityred}
\nonumber\rho^n_{AR}&=&\mu^2\Big[\cos^4r\proj{00}{00}+2\sin^2r\,\cos^2r\proj{01}{01}\\*
\nonumber&&+\sin^4r\proj{02}{02}\Big]+\mu\cos^3r\left(\frac{\beta^*-\gamma^*}{\sqrt2}\proj{00}{11}\right.\\*
\nonumber&&\left.+\frac{\beta-\gamma}{\sqrt2}\proj{11}{00}\right)+(1-\mu^2)\Big[\cos^2r\proj{11}{11}\\*
&&+\sin^2r\proj{12}{12}\Big]
\end{eqnarray}
we can readily compute the partial transpose $\sigma^n=(\rho^n_{AR})^{pT}$
\begin{eqnarray}\label{ptrans}
\nonumber\sigma^n&=&\mu^2\Big[\cos^4r\proj{00}{00}+2\sin^2r\,\cos^2r\proj{01}{01}\\*
\nonumber&&+\sin^4r\proj{02}{02}\Big]+\mu\cos^3r\left(\frac{\beta^*-\gamma^*}{\sqrt2}\proj{01}{10}\right.\\*
\nonumber&&\left.+\frac{\beta-\gamma}{\sqrt2}\proj{10}{01}\right)+(1-\mu^2)\Big[\cos^2r\proj{11}{11}\\*
&&+\sin^2r\proj{12}{12}\Big]
\end{eqnarray}
whose eigenvalues are
\begin{eqnarray}
\nonumber\lambda_1&=&\mu^2\cos^4r\\*
\nonumber\lambda_2&=&\mu^2\sin^4r\\*
\nonumber\lambda_3&=&(1-\mu^2)\cos^2r\\*
\nonumber\lambda_4&=&(1-\mu^2)\sin^2r\\*
\nonumber\lambda_{5,6}&\!=&\!\cos^2r\left(\mu^2\sin^2r\pm\mu\sqrt{\mu^2\sin^4r+\cos^2r\frac{|\beta-\gamma|^2}{2}}\right)
\end{eqnarray}
all the eigenvalues are non-negative except $\lambda_6\le0$. The negativity \eqref{negativity} is, in this case,
\begin{equation}\label{negativitymode}
\mathcal{N}=2\cos^2r\left|\mu^2\sin^2r-\mu\sqrt{\mu^2\sin^4r+\cos^2r\frac{|\beta-\gamma|^2}{2}}\right|
\end{equation}
which depends on the proportion of singlet \mbox{$|\beta-\gamma|/\sqrt2$} of the $\ket{11}$ component in the state \eqref{genstate}. When there is no singlet component ($\beta=\gamma$) the negativity is zero. Indeed in the limit $a\rightarrow0$ (Minkowskian limit)
\begin{equation}\label{negativitymode0}
\mathcal{N}_0=\sqrt2\left|\mu\right|\left|\beta-\gamma\right|
\end{equation}
That shows that the maximally entangled Minkowski occupation number state (Negativity $=1$) arises tracing over total spin when the starting state is
\begin{equation}\label{modesmaxen}
\ket{\Psi}=\frac{1}{\sqrt2}\ket{0,0}\pm\frac12\big[\ket{\uparrow,\downarrow}-\ket{\downarrow,\uparrow}\big]
\end{equation}
or, in the occupation number-total spin bases
\begin{equation}\label{modesmaxen2}
\ket{\Psi}=\frac{1}{\sqrt2}\Big[\ket{00}\ket{S}\pm\ket{11}\ket{S}\big]
\end{equation}
That means that, for occupation number entanglement, the only way to have an entangled state of the bipartite vacuum $\ket{00}$ and the one particle state $\ket{11}$ of a Dirac field is through the singlet component of total spin for the $\ket{11}$ component.

On the contrary, the state
\begin{equation}
\ket{\Psi}=\frac{1}{\sqrt2}\Big[\ket{00}\ket{S}\pm\ket{11}\ket{T_{0,\pm}}\big]
\end{equation}
will become separable after tracing over total spin due to the orthonormality of the basis \eqref{e1}, \eqref{e2}.

We have established that the Minkowski maximally entangled state for occupation number arises after tracing over total spin in a state as \eqref{modesmaxen}. Now we will compute the limit of the negativity when the acceleration goes to $\infty$ in order to see its Unruh decoherence and to compare it with the results for occupation number entanglement from \cite{AlsingSchul}.

Taking $a\rightarrow\infty \Rightarrow r\rightarrow\pi/4$ in \eqref{negativitymode}
\begin{equation}\label{negativitymodeinfty}
\mathcal{N}_\infty=\frac12\left|\mu^2-\sqrt{\mu^4+\mu^2|\beta-\gamma|^2}\right|
\end{equation}
Therefore, for the maximally Minkowski entangled state we have $\mu=1/\sqrt{2}$, $|\beta-\gamma|=1$ and the negativity in the limit is
\begin{equation}\label{negalimi}
\mathcal{N}_\infty=\frac{\sqrt3-1}{4}
\end{equation}
This result shows that when we are reducing the total spin information, looking at the occupation number entanglement alone, we see that it is more degraded by Unruh effect than when we considered spin Bell states in the previous section. More importantly, the occupation number entanglement is more degraded than in \cite{AlsingSchul}, where the spin structure of the modes was considered nonexistent. This happens because considering spin structure of each mode, occupation number 2 is allowed. Hence, Pauli exclusion principle protection of the entanglement is weaker than in \cite{AlsingSchul} where the spin is not considered. The negativity dependence on the acceleration is shown in fig. 3
\begin{figure}\label{fig5}
\includegraphics[width=.45\textwidth]{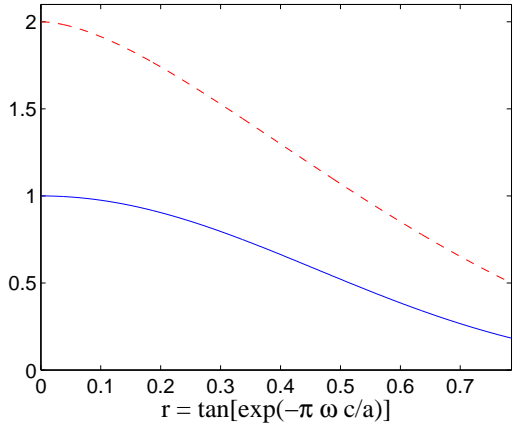}
\caption{Negativity (blue solid line) and mutual information (red dashed line) as a function of the acceleration of Rob when R and A share an occupation number maximally entangled state \eqref{modesmaxen} in Minkowski coordinates after tracing over total spin}
\end{figure}

We can also compute the mutual information for the state \eqref{densityred} as we did in the rest of the cases. Its analytical expression is quite long and has no special interest, but we can see the dependence of $I_{AR}$ for the Minkowski maximally entangled state \eqref{modesmaxen} with the acceleration in fig. 3, obtaining that $I_{AR}^0=2$ and $I_{AR}^\infty=1/2$

\section{Conclusions and comments}\label{sec7}

It is known \cite{Alicefalls,AlsingSchul} that Unruh decoherence degrades entanglement of occupation number states of fields. Here we have shown a richer casuistic that appears when we take into account that each Dirac mode has spin structure. This fact enables us to study interesting effects (such as Unruh decoherence for spin Bell states) and develop new procedures to erase spin information from the system in order to study occupation number entanglement.

Along this work we have analyzed how a maximally entangled spin Bell state losses entanglement when one of the partners accelerates. We have seen that, while in Minkowski coordinates Alice and Rob have qubits, when Rob accelerates the system becomes a non-pure state of a qubit for Alice and a qutrit for Rob. In this case spin entanglement for a Dirac field is degraded when Rob accelerates. However some degree of entanglement survives even at the limit $a\rightarrow\infty$.

A first analog to the well studied state $(1/\sqrt{2})(\ket{00}+\ket{11})$ but including spin could be, for instance, $(1/\sqrt{2})(\ket{00}+\ket{\uparrow\downarrow})$. This state, unlike the deceivingly similar spin Bell states, becomes a qubit$\times$qu4it when Rob accelerates. Nevertheless, distillable entanglement degrades in the same way as for spin Bell states.

We have also introduced a procedure to consistently erase spin information from our setting preserving the occupation number information. We have done it by tracing over total spin. The maximally entangled occupation number state is obtained from the total spin singlet \eqref{modesmaxen} after tracing over total spin. Finally we have shown that its entanglement and mutual information is more degraded than in \cite{AlsingSchul} where the spin structure of Dirac modes was neglected. A reasonable physical argument for this result is that, in our setting, occupation number 2 is allowed for the Dirac field modes, and hence, there is a broader margin for entanglement degradation by Unruh effect.

The thermal noise \eqref{Unruh} is obtained when dealing with a two dimensional space-time and massless fields. Mass gap and transverse degrees of freedom modify the counting statistics that is no longer given by thermal noise, but replaced by the so-called Rindler noise \cite{Takagi}, that 
 the the space-time dimension. In this work we were concerned with the specific issues associated to the spin degree of freedom, so the restriction to massless fields in two dimensional space-time adopted here, as well as the single mode approximation, allows a direct comparison with the previous works \cite{Alicefalls,AlsingSchul} which considered massless spinless fields in 2D under those approximations.

As a matter of fact, having more than 2 space-temporal dimensions and having massive fields may introduce relevant effects. Allowing the possibility of having momentum off the acceleration direction and having massive fields we would obtain a spread of Minkowski modes over Rindler frequencies \cite{Takagi}. If we carry out the single-mode approximation, the spread of Minkowski modes into Rindler modes can be neglected even for higher dimensions \cite{AlsingSchul,Alsingtelep,AlsingMcmhMil}, but if we want to relax such an approximation (as for the discussion in the next paragraph), those effects should be considered in order to account for the entanglement in non-inertial frames. It would be worthwhile to study those effects in future articles.

Another very interesting point that deserves further study is the fact that when we consider more than one populated mode of the complete Minkowski vacuum \eqref{vacua} instead of the single mode approximation, the margin for Unruh degradation increases as we could have in principle a larger number of levels that can be excited by the Thermal/Rindler noise. One could think that these cases would be quite similar to the bosonic case \cite{Alicefalls} where the margin for Unruh decoherence is so broad that no entanglement survives at the limit of $a\rightarrow\infty$. Something similar would apply as well if we relax the single mode approximation allowing a small spread in both Rob's detector and populated levels such that we consider a continuum of accessible levels. These topics will be inspiration for future works.

\section{Acknowledgements}

The authors thank C. Sabin and J. Garcia-Ripoll for the useful discussions during the elaboration of this paper.

This work has been partially supported by the Spanish MCIN Project FIS2008-05705/FIS. E. M-M is partially supported by the CSIC JAE-PREDOC2007 Grant.

We are in debt to an anonymous referee, that has pointed out very interesting questions which helped us substantially improve this article.

\appendix

\section{Some Minkowski operators expressed in Rindler's region I}\label{appen}

To help with the calculations of the density matrix associated with \eqref{genstate} it is useful to compute firstly the trace over IV on all the operators that compose \eqref{Minkowdens}. Using equation \eqref{vacuum} we have
\begin{eqnarray}
\nonumber\proj{0,0}{0,0}&=&\ket{0_A}\left[\cos^2 r\,\biket{0}{0}+e^{i\phi}\sin r\,\cos r\right.\\*
\nonumber &&\times\left(\biket{\uparrow}{\downarrow}+\biket{\downarrow}{\uparrow}\right)+e^{2i\phi}\sin^2 r\\*
&&\left.\times\biket{\pa}{\pa}\right]\otimes \text{H.c.}
\end{eqnarray}
tracing over IV:
\begin{eqnarray}\label{trazapa1}
\nonumber\tr_{IV}\proj{0,0}{0,0}&=&\cos^4 r\,\proj{0,0}{0,0}\\*
\nonumber&&+\sin^2r\,\cos^2r\left(\proj{0,\uparrow}{0,\uparrow}+\proj{0,\downarrow}{0,\downarrow}\right)\\*
&&+\sin^4r\,\proj{0,\pa}{0,\pa}
\end{eqnarray}
where notation is different in each side of the equality: bras and kets in l.h.s. are referred to Minkowski coordinates for Alice and Rob $\ket{s,s'}=\ket{s_A}\ket{s'_R}$ and in the r.h.s. they are referred to Alice's mode in Minkowski coordinates and Rob's mode in Rindler's region I $\ket{s,s'}=\ket{s_A}\ket{s'_R}_I$.

In the same way, using expressions \eqref{vacuum},\eqref{onepart2} we have
\begin{eqnarray}
\nonumber\proj{0,0}{s,s'}&=&\ket{0_A}\left[\cos^2 r\,\biket{0}{0}+e^{i\phi}\sin r\,\cos r\right.\\*
\nonumber &&\times\left(\biket{\uparrow}{\downarrow}+\biket{\downarrow}{\uparrow}\right)+e^{2i\phi}\sin^2 r\\*
\nonumber &&\left.\times\biket{\pa}{\pa}\right]\bra{s}\left(\cos r\,{_I\!\!\bra{s'}}\,{_{IV}\!\!\bra{0}}\right.\\*
&& \left.+ \varepsilon'\, e^{-i\phi}\sin r\,{_I\!\!\bra{\pa}}\,{_{IV}\!\!\bra{s'}}\right)
\end{eqnarray}
where $\varepsilon'=1$ if $s=\uparrow$ and $\varepsilon'=-1$ if $s=\downarrow$. Now, tracing over IV
\begin{eqnarray}\label{trazapa2}
\nonumber\tr_{IV}\proj{0,0}{s,s'}&=&\cos^3 r\,\proj{0,0}{s,s'}+\sin^2r\,\cos r\\*
\nonumber&&\times\left(\delta_{s'\uparrow}\proj{0,\downarrow}{s,\pa}-\delta_{s'\downarrow}\proj{0,\uparrow}{s,\pa}\right)\\*
\end{eqnarray}
notation here is the same than in \eqref{trazapa1}.

Again, using expression \eqref{onepart2} we get
\begin{eqnarray}
\nonumber\proj{s_1,s_2}{s_3,s_4}&=&\ket{s_{1}}\left[\cos r\,\biket{s_2}{0}+\varepsilon_2e^{-i\phi}\sin r\right.\\*
\nonumber&&\left.\times\biket{\pa}{s_2}\right]\bra{s_{3}}\left[\cos r\,{_I\!\!\bra{s_4}}{\,_{IV}\!\!\bra{0}}\right.\\*
&&\left.+\varepsilon_4e^{-i\phi}\sin r\,{_I\!\!\bra{\pa}}{\,_{IV}\!\!\bra{s_4}}\right]
\end{eqnarray}
and tracing over IV gives
\begin{eqnarray}\label{trazapa3}
\nonumber\tr_{IV}\proj{s_1,s_2}{s_3,s_4}&=&\cos^2 r\,\proj{s_1,s_2}{s_3,s_4}\\*
\nonumber&&+\delta_{s_2s_4}\sin^2r\proj{s_1,\pa}{s_3,\pa}\\*
\end{eqnarray}
again, notation here is the same than in \eqref{trazapa1}.

\bibliographystyle{apsrev}
\end{document}